\documentclass[smallextended]{svjour3}
\smartqed
\usepackage{amsmath}
\usepackage{amsfonts}
\usepackage{amssymb}
\usepackage{graphicx}
\usepackage{mathptmx}
\RequirePackage{flushend}
\RequirePackage[numbers,sort&compress]{natbib}
\RequirePackage[colorlinks,citecolor=blue,urlcolor=blue,linkcolor=blue]{hyperref}
\usepackage{subfig}

\journalname{Gen Relativ Gravit}
\begin{document}

\title{Density perturbation in the models reconstructed from jerk parameter}

\author{Srijita Sinha \and Narayan Banerjee}

\authorrunning{S. Sinha \and N. Banerjee}

\institute{Srijita Sinha \at
              IISER Kolkata, Mohanpur Campus, Mohanpur, Nadia, 741246, India\\
                          \email{ss13ip012@iiserkol.ac.in}           
           \and
           Narayan Banerjee \at
              IISER Kolkata, Mohanpur Campus, Mohanpur, Nadia, 741246, India \\
              \email{narayan@iiserkol.ac.in} 
}


\maketitle

\begin{abstract}
 The present work deals with the late time evolution of the linear density contrast in the dark energy models reconstructed from the jerk parameter. It is found that the non-interacting models are favoured compared to the models where an interaction is allowed in the dark sector. 


\end{abstract}

\section{Introduction}
\label{intro}
Ever since the stunning results from the redshift luminosity of supernovae surveys in the final decade of the last millennium\cite{riess, perl, schmidt}, it is believed that our universe is expanding with an acceleration for several Giga years. Certainly there have been counter arguments, but no alternative explanation for the dimming of the supernovae could survive the observational evidences. For a very recent review, we refer to the work of Haridasu {\it et al.}\cite{haridasu} and also Rubin and Hayden\cite{rubin}. There must be an agent which enables the universe to overcome the attractive interaction--- gravity, and speed up like this. The debate over the fittest candidate that can drive the acceleration, the so-called dark energy (DE), is yet to be settled. The cosmological constant $\Lambda$ is definitely a very strong candidate, but it has the problem of a huge discrepancy between the observationally required value and the theoretically predicted one. However, it should be appreciated that a $\Lambda$ with a proper value still does the best against observations without essentially jeopardising most of the success of the standard cosmology. So any model leading to an accelerated expansion is quite often framed in such a way so that it actually mimics a $\Lambda$CDM at the present epoch. A quintessence field, which is essentially a scalar field endowed with a potential is another popular candidate\cite{durrive}. Detailed reviews for various aspects of a cosmological constant and quintessence models exist from the very early days of the idea of a dark energy\cite{paddy, varun, sami,carroll, ratra, martin}. The list is far from being exhaustive. The story so far is summarised in a recent work by Brax\cite{brax}. \\

In the absence of a universally accepted model for the accelerated expansion, attempts have also been there to ``reconstruct'' a dark energy model. The idea is to guess a model of evolution that explains the observations and find the distribution of matter that can give rise to that\cite{em}. As there is no dearth of observational data now, various improvisations of the methods of reconstruction have been suggested, making use of the data efficiently. For a scalar field model, Starobinsky showed that one can exploit the data on density perturbation to reconstruct the scalar potential\cite{staro}. Huterer and Turner, on the other hand, utilized the distance measurement data for the same purpose\cite{huterer1, huterer2}. \\

Reconstruction of dark energy models normally start with some physical quantity, like the potential of the quintessence field, or the equation of state parameter $w$ which is the ratio of the pressure and the energy density of the corresponding matter\cite{varun, saini1, gerke,gong,holsclaw, polarski, linder, anjan1, anjan2, hu, slp1, slp2, czp}. \\

There is a new trend in reconstruction of dark energy models where one completely ignores the physical quantities and bank on kinematical parameters. With the assumption of spatial isotropy and homogeneity, the spacetime geometry is determined by the scale factor $a$. The derivatives of $a$ with respect to the cosmic time of various orders will yield the kinematical quantities. The first few of them are the Hubble parameter $H = \frac{\dot{a}}{a}$, the deceleration parameter $q = - \frac{a\ddot{a}}{{\dot{a}}^2}$, the jerk parameter $j = - \frac{1}{a^3} \frac{d^3 a}{dt^3}$ and so on. The basic idea is to write an ansatz for some kinematical quantity which involves some parameters and then an estimation of the parameters using observational data. In a way, this is an attempt to construct the model through cosmography, where one builds up the model from observables rather than modelling from a theory. Cosmography may be related to a particular data set. For example, the baryon acoustic oscillation\cite{alcaniz}, supernovae data\cite{cattoen}, observations at high redshift\cite{vitagliano} to name a few. Cosmographic methods without using standard candles and standard rulers are discussed by Xia {\it et al.}\cite{xia}. Cosmography, using a Markov Chain method, has been discussed by by Capozziello, Lazcoz and Salzano\cite{salzano}. For a very recent account of cosmography, one can see \cite{li}. \\

Hubble parameter is the oldest observational quantity in cosmology and it was found to have an evolution. So the natural choice for a reconstruction through a kinematical quantity has been the next higher order derivative, namely the deceleration parameter $q$. Attempts to build up a dark energy model with $q$ as the starting point have been made by Gong and Wang\cite{gong} and by Ting \textit{et al.}\cite{wang}. The deceleration parameter $q$ at various epoch can be estimated today with the help of observational quantities like luminosity of supernovae and the corresponding redshift, and is found to be evolving as well. For a recent work on this estimation, we refer to the work of vanPutten\cite{vanPutten}. Hence the next higher order derivative of the scale factor, the jerk parameter $j$, is of importance now and should play a significant role in the game of reconstruction through kinematical quantities. Some work in this direction has been initiated by Luongo\cite{luongo} and by Rapetti\cite{rapetti}. Zhai \textit{et al.}\cite{zz}, starting from a parametric ansatz for $j$ such that the present (at $z=0$) value of $j$ is -1, reconstructed quite a few dark energy models. The motivation behind the choice of $j=-1$ at $z=0$ is the fact that it mimics the standard $\Lambda$CDM model of the present acceleration of the universe. That the jerk parameter should be instrumental for the reconstruction through kinematical quantities was categorically indicated a long time back by Alam \textit{et al.}\cite{alam}.\\

In a very recent work, Mukherjee and Banerjee\cite{ankan1} relaxed the requirement that $j(z=0)=-1$, and reconstructed several dark energy models. This is more general in the sense that the present value of $j$ is not controlled by hand. This  work also involves diverse data sets for the estimation of the model parameters as opposed to the work by Zhai \textit{et al.}\cite{zz} where only Observational Hubble Data (OHD) and the Union 2.1 Supernovae data were employed. In a later work, Mukherjee and Banerjee\cite{ankan2} looked at the possible interaction between the cold dark matter and the dark energy, with the help of a model reconstructed through the jerk parameter, with an assumption that the jerk is varying very slowly, and can be approximated as a constant. \\

The reconstructed models, with a good choice of parameters, can fit well with various observational data. But in order to provide a useful description of the evolution, a model should also be able to describe certain other things. One crucial aspect is certainly a consistency with a growing mode of density perturbations, without which the formation of structures cannot be explained. The motivation of the present work is to investigate the density perturbations and to check whether the models reconstructed via $j$ can give rise to a growing mode of such perturbations. There is hardly any work on the matter perturbations in reconstructed models, except perhaps that of Hikage, Koyama and Heavens\cite{kage}, where the perturbation of a model reconstructed from Baryon Acoustic Oscillation (BAO) data has been discussed. There is, however, some work on the reconstruction of perturbation itself\cite{gonzal1, gonzal2, hunt, alam1}.\\ 

The present work deals with the density perturbation of models reconstructed from an ansatz on the jerk parameter. For a varying jerk, we pick up the models from reference \cite{ankan1}, as that is more general. We also consider interacting models, where the jerk is very slowly varying, as in reference \cite{ankan2}. \\

It should be realized that the perturbations of the model based on Einstein equations and that of a kinematically reconstructed model could well be different even for the same energy budget of the universe. This is for the simple reason that in the former, one has more independent equations and thus contributions to perturbation from other sectors, such as the velocity perturbation could be manifest. \\

The paper is organized as follows. Section 2 discusses about the background, where the only theoretical ansatz is spatial isotropy and homogeneity of the spacetime resulting in an FRW metric. The next section deals with standard distribution of matter, which will be used only as an analogy and not for the actual perturbation. Section 4 deals with a scenario where the jerk parameter is practically a constant but an interaction in the matter sector is allowed. In section 5, a varying jerk is considered where no interaction in the matter sector is allowed. In the last section, we summarize and discuss the results.

\section{Background} \label{sec2}

A spatially flat, homogeneous and isotropic universe is given by the metric
\begin{equation}\label{metric}
ds^2=-d t ^2+a^2(t)\gamma_{ij} d x^i dx^j,
\end{equation}
where $\gamma^i_{j}=\delta^i_j$ is the metric in the constant time hypersurface and $a(t)$ is the scale factor. The kinematic quantities of our interest are \\
\begin{enumerate}
\item[(i)] the fractional first order time derivative of the scale factor $a$, the well known Hubble parameter, $H=\frac{\dot{a}}{a}$; \\
\item[(ii)] the second order derivative of $a$, defined in a dimensionless way, the  deceleration parameter, $q=-\frac{\ddot{a}/a}{\dot{a}^2/a^2}$; \\
\item[(iii)] the third order derivative of $a$, again defined in a dimensionless way, the jerk parameter $j(t)=-\frac{1}{a H^3}\left(\frac{d^3 a}{d t^3}\right)$. \\
\end{enumerate}

We pick up the convention in which $j$ is defined with a negative sign so that we can make use of the results from references \cite{zz,ankan1,ankan2} without any modification. In terms of redshift, $z=\frac{a_0}{a}-1$, ($a_0$ is the present value of the scale factor, taken to be unity throughout the calculation), the jerk parameter takes the form
\begin{equation} \label{jerk}
j(z) = -1 + \left(1+z \right) \frac{\left(h^2\right)'}{h^2} -\frac{1}{2}\left(1+z \right)^2 \frac{\left(h^2\right)''}{h^2},
\end{equation}
where $h(z)=\frac{H(z)}{H_0}$, $H_0$ being the present value of the Hubble parameter and prime denotes a differentiation  with respect to redshift, $z$.

\section{Standard distribution of matter}\label{stanmat}

Although we shall be working with equation (\ref{jerk}), pretending that we do not know anything about the matter distribution in the universe, we should be able to identify the terms with the corresponding standard matter distribution consisting of a dark matter and a dark energy at some stage. \\

The energy-momentum tensor of a perfect fluid distribution is
\begin{equation} \label{stress}
T^{(m)}{}_{\mu \nu}= (p_m + \rho_m)u_\mu u_\nu + p_m g_{\mu \nu},
\end{equation}
where $\rho_m$ is the fluid density, $p_m$ is the pressure and $u_\mu$ is the comoving $4$-velocity, $u_{\mu} = (1,0,0,0)$. For a cold dark matter (CDM) $p_m=0$.\\

The contribution to the density and pressure from the dark energy are $\rho_{de}$ and $p_{de}$ respectively. The dark energy is also assumed to mimic a fluid distribution, and the corresponding energy momentum tensor is the same as equation (\ref{stress}) with $\rho_{m}$ and $p_m$ being replaced by $\rho_{de}$ and $p_{de}$ respectively. The equation of state (EoS) parameter of the dark energy component is $w_{de} = \frac{p_{de}}{\rho_{de}}$.\\

If an interaction between the dark matter and the dark energy is allowed, they do not conserve individually, and one can write the rate of transfer of energy in terms of an interaction term $\eta$ so that 
\begin{eqnarray} 
\dot{\rho}_m + 3 H \rho_m &=& \eta , \label{econ1}\\
\dot{\rho}_{de} + 3 H \left(1+w_{de}\right) \rho_{de} &=& -\eta \label{econ2}.
\end{eqnarray}
Thus the above two equations combine to give the total conservation equation, which would follow from the standard Einstein equations. In the absence of any interaction, both the components conserve individually, indicating $\eta=0$.

\section{A slowly varying jerk} \label{sec3}

If we now consider the jerk to be a very slowly varying function of $z$ so that one can consider it to be a constant for the purpose of integration\cite{ankan2}, the equation (\ref{jerk}) can be integrated to yield 

\begin{equation} \label{hubble1}
h^2(z) =  A \left(1+z\right)^{\frac{3+\sqrt{9-8(1+j)}}{2}} +  \left(1-A\right) \left(1+z\right)^{\frac{3-\sqrt{9-8(1+j)}}{2}} .
\end{equation}

Here $A$ is a constant of integration. From definition, $h^2(z=0)$ has to be unity, so that the second constant of integration is chosen to be $(1-A)$ in order to satisfy the condition. From equation (\ref{hubble1}), one can easily see that for $j=-1$, which corresponds to a $\Lambda$CDM model, the first term redshifts as a pressureless fluid and the second term corresponds to a constant. With this identification, the first term is easily picked up as the matter density parameter $\Omega_m$ and the second term as the dark energy density parameter $\Omega_{de}$ which reduces to a cosmological constant for $j=-1$. So one actually recovers the $G^{0}_{0}$ component of Einstein equations in terms of the dimensionless quantities as,
\begin{equation} \label{hubble2}
h^2 = \Omega_m + \Omega_{de},
\end{equation}
where 
\begin{eqnarray}
\Omega_m &=& A \left( 1+z \right)^{\frac{3+\sqrt{9-8(1+j)}}{2}}, \label{omega1}\\
\mbox{and} \hspace{2mm} \Omega_{de}&=&\left(1-A\right) \left(1+z\right)^{\frac{3-\sqrt{9-8(1+j)}}{2}}. \label{omega2}
\end{eqnarray}
For values of $j$ other that $-1$, the second term is clearly an evolving dark energy rather than a constant i.e., $\Lambda$. The sector identified to be the dark matter does not redshift as $(1+z)^3$ in this case. However one can still identify that with the standard pressureless matter but has to allow an interaction amongst the dark sector as shown in reference \cite{ankan2}.  The interaction, $\eta$ between the two components can be expressed in terms of $z$ as
\begin{equation} \label{int}
\eta(z) = A \rho_c \left(\frac{3-\sqrt{9-8(1+j)}}{2}\right) \left(1+z\right)^\frac{3+\sqrt{9-8(1+j)}}{2} H(z).
\end{equation}
The EoS of dark energy in terms of $z$ is obtained as
\begin{equation}\label{eos}
\begin{split}
w_{de}(z)= - \left(\frac{3+\sqrt{9-8(1+j)}}{6}\right)& \\
 -\left(\frac{A}{1-A}\right)& \left(\frac{3-\sqrt{9-8(1+j)}}{6}\right) \left(1+z\right)^{\sqrt{9-8(1+j)}} .
\end{split}
\end{equation}

Now small perturbations of densities and interaction are considered in the form $\bar{\rho}_m= \rho_m + \delta \rho_m$, $\bar{\rho}_{de}=\rho_{de}+\delta \rho_{de}$ and $\bar{\eta}= \eta +\delta \eta$ and the resulting metric perturbation as , $\bar{H}=H+\delta H$ in equations (\ref{econ1}), (\ref{econ2}) and (\ref{hubble2}). Expanding upto the first order the equations are obtained respectively as
\begin{eqnarray}
-H \left(1+z\right) \delta \rho_m' + 3 H \delta \rho_m + 3 \delta H \rho_m &=& \delta \eta , \label{pert1}\\
-H \left(1+z\right) \delta \rho_{de}' + 3 H \left(1+w_{de}\right) \delta \rho_{de} + 3 \delta H \left(1+w_{de}\right) \rho_{de} &=& -\delta \eta , \\
2 H \delta H = \delta \rho_m& + &\delta \rho_{de} \label{pert2}.
\end{eqnarray}
Using equations (\ref{int}) and (\ref{eos}) in (\ref{pert1})-(\ref{pert2}), a second order differential equation for the density contrast of the dark matter $\delta=\frac{\delta \rho_m}{\rho_m}$ is obtained as 
\begin{equation} \label{eq1}
\begin{split}
&2 \left( 1+z\right)^2 H^2 \rho_m \delta''(z)\\
-&\frac{\delta'(z)}{A \beta  \rho_{c} \left( 1+z\right)^{\alpha }-3 \rho_{m}}\left(\left( 1+z\right) \left(-4 \left( 1+z\right) H \rho_{m} H' \left(A \beta  \rho_{c} \left( 1+z\right)^{\alpha }-3 \rho_{m}\right)\right.\right.\\
+&\left.\left.2 H^2 \left(\rho_{m} \left(3 A \beta  \rho_{c} \left( 1+z\right)^{\alpha }w_{de}+A \left(\alpha +5\right) \beta  \rho_{c} \left( 1+z\right)^{\alpha }+3 \left( 1+z\right) \rho_{m}'\right)\right.\right.\right.\\
-&\left.\left.\left.2 A \beta  \rho_{c} \left( 1+z\right)^{\alpha +1} \rho_{m}'-3 \rho_{m}^2 \left(3 w_{de}+5\right)\right)\right.\right.\\
+&\left.\left.3 \rho_{m} \left(\rho_{de} \left(w_{de}+1\right)+\rho_{m}\right) \left(A \beta  \rho_{c} \left( 1+z\right)^{\alpha }-3 \rho_{m}\right)\right)\right)\\
-& \frac{\delta(z)}{A \beta  \rho_c \left( 1+z\right)^{\alpha }-3 \rho_m}\left(-4 \left( 1+z\right) H H' \left(\left( 1+z\right) \rho_m'-3 \rho_m\right)\right.\\
&\left. \left(A \beta  \rho_c \left( 1+z\right)^{\alpha }-3 \rho_m\right)-2 H^2 \left(3 \rho_m \left(3 w_{de}\right.\right.\right.\\
&\left.\left.\left. \left(A \beta  \rho_c \left( 1+z\right)^{\alpha }+\left( 1+z\right) \rho_m'\right)\right.\right.\right.\\
+& \left.\left.\left.3 A \beta  \rho_c \left( 1+z\right)^{\alpha }+A \alpha  \beta  \rho_c \left( 1+z\right)^{\alpha }-\left( 1+z\right)^2\rho_m''+2 \left( 1+z\right) \rho_m'\right)\right. \right.\\
+& \left. \left.\left( 1+z\right) \left( A \beta  \rho_c \left( 1+z\right)^{\alpha +1} \rho_m''\right.\right.\right.\\
&\left.\left.\left.-A \beta  \rho_c \left( 1+z\right)^{\alpha } \left(\alpha +3 w_{de}+5\right) \rho_m'+3 \left( 1+z\right) \rho_m'^2\right)\right. \right.\\
-& \left. \left. 27 \rho_m^2 \left(w_{de}+1\right)\right)+3 \left(A \beta  \rho_c \left( 1+z\right)^{\alpha }-3 \rho_m(z)\right) \left(-\rho_m\right.\right.\\
&\left.\left. \left(-A \beta  \rho_c \left( 1+z\right)^{\alpha } w_{de}-\left( 1+z\right) \rho_m'+3 \rho_m \left(w_{de}+1\right)\right)\right.\right.\\
-&\left.\left.\rho_{de} \left(w_{de}+1\right) \left(3 \rho_m-\left( 1+z\right) \rho_m'\right)\right)\right)=0
\end{split}
\end{equation}
where $\alpha=\frac{3+\sqrt{9-8(1+j)}}{2}$ and $\beta=\frac{3-\sqrt{9-8(1+j)}}{2}$. \\

\begin{figure}[!htbp]
  \centering
\includegraphics[width=1\textwidth]{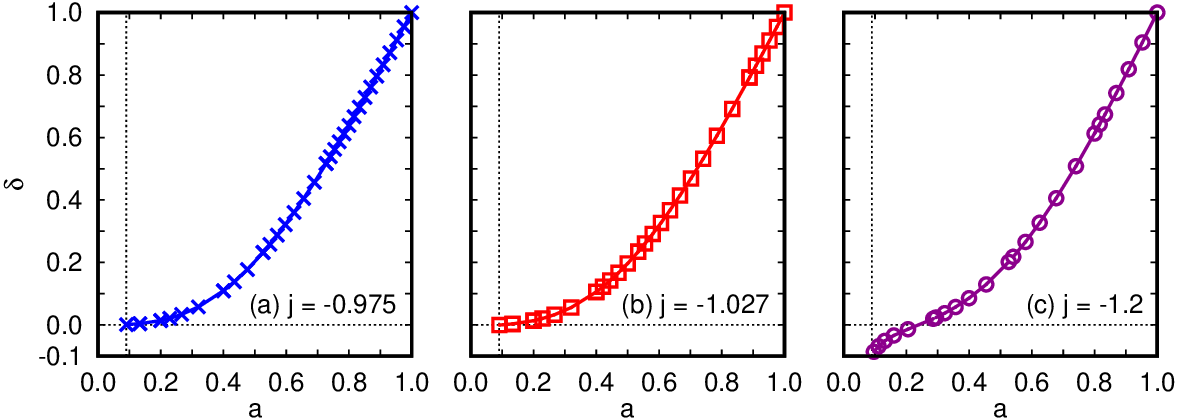}
\caption{Plot of $\delta$ against $a$ for $j=-0.975$ , $-1.027$ and $-1.20$. The vertical dotted line corresponds to $x=0.09$ and the horizontal dotted line corresponds to the zero crossing line.}\label{fig1}
\includegraphics[width=0.4\textwidth]{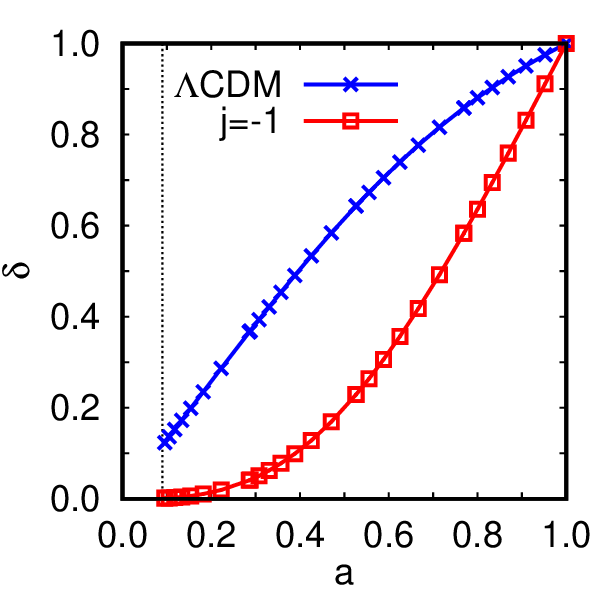}
\caption{Plot of $\delta$ against $a$ for $\Lambda$CDM and $j=-1$. The vertical dotted line corresponds to $x=0.09$.}\label{fig2}
\end{figure}

The differential equation (\ref{eq1}) gives the evolution of the density contrast $\delta$ with $z$. This equation is solved numerically from $z=1100$ to the present epoch $z=0$ with standard initial conditions as given by Cembranos  \textit{et al.}\cite{jfbd} and Mehrabi  \textit{et al.}\cite{mehrabi} like $\delta(z=1100)=0.001$ and $\delta'(z=1100)=0.$ The value of the constant $A$, which is actually $\Omega_{m0}$ is taken as $0.286$\cite{ankan2}. It deserves mention that the value is not taken from any particular observation, it is rather the best fit value as found by the statistical analysis given in reference \cite{ankan2}. In order to get a qualitative picture of the perturbation, we scale $H_0$ to unity. The estimates are in gravitational units, where $G=1$. The evolution of $\delta$ is investigated for three different values of $j$ namely, $-1.027$, $-0.975$ and $-1.2$. The best fit value is $j=-1.027$, and the other two are  roughly the two extremes of the $2 \sigma$ confidence region as given in \cite{ankan2}. Although the initial conditions are taken for $z=1100$, we show the plots of $\delta$ between $a=1$ (i.e.\ $z=0$) and $a=0.09$ (i.e.\ $z=10$) so as to have a closer look into the late time behaviour. The plots for the whole domain will give a much poorer resolution. The qualitative behaviour is shown in figure (\ref{fig1}). In order to have tractable plots, we have normalised the value of $\delta$ by that at $a=1$. It is quite clearly seen from the plots that close to $a=1$, all the plots are linear.

One can see that for less that the best fit value of $j$, the density contrast changes sign with the evolution. However, if $j> - 1.027$, the best fit value, the matter perturbation has a monotonic growing mode. Thus even within the $2\sigma$ confidence level, the interacting model has a problem.

In figure (\ref{fig2}) we show the plots of the standard $\Lambda$CDM model. Vale and Lemos\cite{vale} gave the evolution equation of $\delta$ for $\Lambda$CDM as \\
\begin{equation}\label{lcdm}
(z+1)^2 \delta_{\Lambda}^{\prime \prime}(z) + (z+1) \left(-1+\frac{(z+1) H^{\prime}(z)}{H(z)}\right) \delta_{\Lambda}^{\prime}(z)+\left(\frac{\Lambda }{2 H(z)^2}-\frac{3}{2}\right) \delta_{\Lambda}(z)=0.
\end{equation}
We numerically integrate this with the boundary conditions used in the present work and plot the result in figure (\ref{fig2}) in a blue crossed line. The one in a red squared line is the plot of $\delta$ against $a$ from our numerical analysis of equation (\ref{eq1}) with $j=-1$ which is supposed to correspond to a $\Lambda$CDM.  The plots are not really coincident even though the same initial conditions are used to integrate the equations (\ref{eq1}) and (\ref{lcdm}). Both the plots are linear close to $a=1$, but definitely with a different slope. The difference is due to the following reason. The present model has only one equation, namely the equation (\ref{hubble1}), which is used for the perturbation. A standard $\Lambda$CDM model, on the other hand, has an additional equation, that for $\dot{H}$. So the model from the reconstruction of jerk indeed mimics the $\Lambda$CDM kinematically, but they are not really identical. Perturbations of the models strongly indicate that.\\

\section{A varying jerk}
We now relax the requirement of a slowly varying jerk and allow the jerk to be a function of $z$. But the interaction in the dark sector is switched off, such that both DM and DE have their own conservation as in reference \cite{ankan1}.\\

The general form of $j$, as suggested by Mukherjee and Banerjee\cite{ankan1} is
\begin{equation} \label{jerkdef}
j(z)= -1 +j_1\frac{f(z)}{h^2(z)},
\end{equation}
where $f$ is an analytic function of $z$ and $j_1$ is parameter to be determined by observational data. The four different forms of $f(z)$ from \cite{ankan1} are
\begin{eqnarray}
\mbox{Case I:} \hspace{1cm} j(z)&=&-1+j_1\frac{1}{h^2(z)},\\ \label{casei}
\mbox{Case II:}\hspace{1cm} j(z)&=&-1+j_1\frac{\left(1+z\right)}{h^2(z)},\\
\mbox{Case III:}\hspace{1cm} j(z)&=& -1+j_1\frac{\left(1+z\right)^2}{h^2(z)},\\
\mbox{Case IV:}\hspace{1cm} j(z)&=&-1+j_1\frac{1}{\left(1+z\right)h^2(z)}. \label{caseiv}
\end{eqnarray}

The second order differential equation (\ref{jerk}) can be integrated using these expressions for $j$ to get four different cases as
\begin{eqnarray}
\mbox{Case I:} \hspace{1cm} h^2(z)&=& c_1 \left(1+z\right)^3+c_2 +\frac{2}{3} j_1 \log\left(1+z\right), \label{hubble3}\\
\mbox{Case II:}\hspace{1cm} h^2(z)&=&c_1 \left(1+z\right)^3+c_2 +j_1 \left(1+z\right),\\
\mbox{Case III:}\hspace{1cm} h^2(z)&=& c_1 \left(1+z\right)^3+c_2 +j_1 \left(1+z\right)^2,\\
\mbox{Case IV:}\hspace{1cm} h^2(z)&=&c_1 \left(1+z\right)^3+c_2 +j_1 \frac{1}{2\left(1+z\right)} \label{hubble3a},
\end{eqnarray}
where $c_1$ and $c_2$ are integration constants, which can be evaluated using initial data. The constants $c_1$, $c_2$ and the model parameter $j_1$ are connected by the fact that $h(z=0) = 1$ from its definition. The values of $j_1$ and $ c_1$ are used from \cite{ankan1} as given in table (\ref{tab:title1}).\\

\begin{table}[htbp]
\caption{Values of the constants $c_1$ and $j_1$.} \label{tab:title1} 
\begin{tabular}{lll}
\hline\noalign{\smallskip}
Cases & value of $c_1$ & value of $j_1$ \\ 
\hline\noalign{\smallskip}
Case I & $0.2985$ & $0.078$ \\ 
Case II & $0.299$ & $0.045$ \\ 
Case III & $0.30$ & $0.017$ \\ 
Case IV & $0.298$ & $0.112$ \\ 
\hline\noalign{\smallskip}
\end{tabular}
\end{table}

As the left hand side, of all the equations (\ref{hubble3})-(\ref{hubble3a}), is the square of the Hubble parameter scaled by its present value, it is easy to pick up the first term in each equation as $\Omega_m$, the density parameter of the cold dark matter which does not interact with the other components of matter, as it redshifts as $(1+z)^3$. Also, the constant of integration $c_1$ can thus be identified with $\Omega_{m,0}$, the value of the density parameter at $z=0$. The rest of the right hand side of equations (\ref{hubble3})-(\ref{hubble3a}) is thus picked up as the net $\Omega_{de}$. Thus the evolution of $\Omega_{de}$, in the four cases, will look like 
\begin{eqnarray}
\mbox{Case I:} \hspace{1cm} \Omega_{de}(z)&=& 1-c_1 +\frac{2}{3} j_1 \log\left(1+z\right), \\
\mbox{Case II:}\hspace{1cm} \Omega_{de}(z)&=&1-j_1-c_1 +j_1 \left(1+z\right),\\
\mbox{Case III:}\hspace{1cm} \Omega_{de}(z)&=& 1-j_1-c_1 +j_1 \left(1+z\right)^2,\\
\mbox{Case IV:}\hspace{1cm} \Omega_{de}(z)&=&1-\frac{j_1}{2}-c_1 +j_1 \frac{1}{2\left(1+z\right)}.
\end{eqnarray}

The corresponding expressions for $w_{de}$, the equation of state parameter of the dark energy, are used as given in \cite{ankan1}.
\begin{eqnarray}
\mbox{Case I:} \hspace{1cm} w_{de}(z)&=& -1 + \frac{\frac{2}{9} j_1}{\frac{2}{3} j_1 \log \left(1+z\right)+\left( 1-c_1\right)} , \\
\mbox{Case II:}\hspace{1cm} w_{de}(z)&=&-1  + \frac{\frac{1}{3} j_1\left(1+z\right)}{j_1 \left(1+z\right)+\left( 1-c_1-j_1\right)},\\
\mbox{Case III:}\hspace{1cm} w_{de}(z)&=& -1  + \frac{\frac{2}{3} j_1 \left(1+z\right)^2}{j_1 \left(1+z\right)^2+\left( 1-c_1-j_1\right)},\\
\mbox{Case IV:}\hspace{1cm} w_{de}(z)&=&-1+ \frac{\frac{j_1}{6\left(1+z\right)} }{\frac{-j_1}{2\left(1+z\right)} +\left( 1-c_1+\frac{1}{2}j_1 \right)}.
\end{eqnarray}

Considering a small perturbation, as discussed in the section (\ref{sec3}),  equations (\ref{econ1}), (\ref{econ2}) and (\ref{hubble2}) are combined to obtain the differential equation for the density contrast $\frac{\delta \rho_m}{\rho_m}= \delta$ as
\begin{equation} \label{eq2}
\begin{split}
& 2\left( 1+z\right)^2  H^2 \rho_m^2 \delta''(z) + \left( -\left( 1+z\right)  \left( 3 \rho_m^2 \left( \rho_m +\rho_{de} \left( 1+ w_{de} \right) \right)\right. \right. \\
-&\left.\left. 4 \left( 1+z\right) H H' \rho_m^2 +2 H^2 \rho_m \left(\rho_m  \left( 5+3 w_{de} \right) -\left( 1+z\right) \rho_m' \right) \right) \right) \delta'(z)\\ 
+&\left(4 \left( 1+z\right) H H' \rho_m \left( -3 \rho_m + \left( 1+z\right)\rho_m'\right) \right. \\
- &\left.  3 \rho_m \left( -\rho_{de} \left( 1+ w_{de} \right) \left( 3 \rho_m -\left( 1+z\right) \rho_m'\right) + \rho_m \left(-3 \rho_m \left( 1+ w_{de} \right) + \left( 1+z\right) \rho_m'\right) \right)  \right.\\
+& \left.  2 H^2 \left( 9 \rho_m^2 \left( 1+ w_{de} \right) \right. - \left. \left( 1+z\right)^2 \rho_m'^2 \right. \right. \\
+& \left. \left. \left( 1+z\right) \rho_m \left( - \left( 2+3 w_{de} \right) \rho_m' + \left( 1+z\right) \rho_m'' \right) \right) \right) \delta(z)=0.
\end{split}
\end{equation}
Equations (\ref{hubble3})-(\ref{hubble3a}) explicitly show that the first term redshifts as $(1+z)^{3}$, it is easily picked up as the contribution from the pressureless dark matter which conserves by itself. So we have to fix $\eta = 0$ in equations (\ref{econ1}) and (\ref{econ2}), allowing both the components to have their own conservation. \\

Equation (\ref{eq2}) is the dynamical equation for the density contrast $\delta$ against the redshift $z$. The said equation is solved numerically for each of the four ansatz mentioned in equations (\ref{casei})--(\ref{caseiv}) as the unperturbed background.\\

\begin{figure}[!htbp]
\centering
\includegraphics[width=0.7\textwidth]{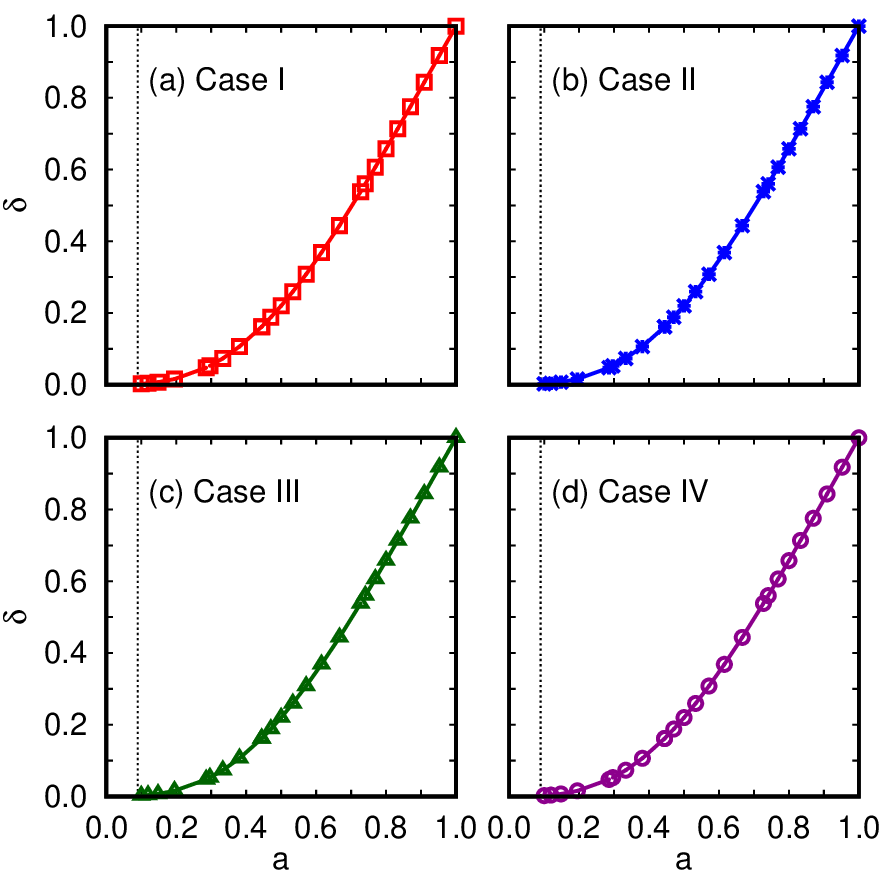}
\caption{Plot of $\delta$ against $a$ for different cases of varying jerk parameter. The vertical dotted line corresponds to $x=0.09$.}\label{fig3}
\end{figure}
Figure (\ref{fig3}) shows the plots of $\delta$ against $a$. All of them appear to be qualitatively similar, and they also possess the required nature of the perturbation, i.e.\ grow with the evolution (increase with $a$). Like the interacting case presented in section (\ref{sec3}), here also the initial conditions are chosen at $z=1100$ as given by Cembranos  \textit{et al.}\cite{jfbd} and Mehrabi  \textit{et al.}\cite{mehrabi}, but the plots are given between $a=0.09$ (corresponding to $z=10$) and $a=1$ for the sake of better resolution.\\

\section{Summary and Discussion}

Albeit there is a proliferation of dark energy models reconstructed from the observational data, their suitability in connection with the structure formation, i.e., the possibility of the growth of density fluctuation, have very rarely been dealt with. The present work is an attempt towards looking at the growth of density perturbation in models reconstructed from the jerk parameter, the kinematical quantity gaining physical relevance recently. The perturbation equations are linearized in the fluctuations. The second order differential equations are solved numerically to plot the density contrast $\delta$ against the redshift $z$. \\

In both the examples of an interacting dark energy and a non-interacting one, the values of the parameters, though identified with some physically relevant quantities like the density parameter, are not taken from observational results, but rather from the best fit values as given in the reconstruction of the respective models. However, the values appear to be not too different from the recent observations like the Planck mission. \\

It appears quite conclusively that the models allowing interaction in the dark sector fail to yield the required behaviour of $\delta$ even within the $2\sigma$ confidence level. The discrepancy is observed close $a=0.09$, i.e., roughly $z=10$. Non-interacting models, however, produce quite a congenial environment for the structure formation, at least qualitatively. The density contrast indeed has growing modes during the later time. So the non-interacting models appear to be favoured so far as the structure formation is concerned.


%

\end{document}